\documentclass[amsmath,twocolumn,showpacs,letterpaper,floatfix]{revtex4}
\usepackage{amssymb}
\usepackage{graphicx}

\newcommand{\rsp}[1]{\hspace{-0.1em}#1\hspace{-0.1em}}

\allowdisplaybreaks[1]

\begin{document}

\title{Griffiths singularities and algebraic order in the exact solution of
an Ising model on a fractal modular network}
\author{Michael Hinczewski}

\affiliation{Feza G\"ursey Research Institute, T\"UB\.ITAK -
Bosphorus University, \c{C}engelk\"oy 34684, Istanbul, Turkey}

\begin{abstract}
  We use an exact renormalization-group transformation to study the
  Ising model on a complex network composed of tightly-knit
  communities nested hierarchically with the fractal scaling recently
  discovered in a variety of real-world networks. Varying the ratio
  $K/J$ of of inter- to intra-community couplings, we obtain an
  unusual phase diagram: at high temperatures or small $K/J$ we have a
  disordered phase with a Griffiths singularity in the free energy,
  due to the presence of rare large clusters, which we analyze through
  the Yang-Lee zeros in the complex magnetic field plane. As the
  temperature is lowered, true long-range order is not seen, but there
  is a transition to algebraic order, where pair correlations have
  power-law decay with distance, reminiscent of the $XY$ model. The
  transition is infinite-order at small $K/J$, and becomes
  second-order above a threshold value $(K/J)_m$. The existence of
  such slowly decaying correlations is unexpected in a fat-tailed
  scale-free network, where correlations longer than nearest-neighbor
  are typically suppressed.
\end{abstract}

\maketitle

\section{Introduction}

Many real-world networks have modular
structure~\cite{GirvanNewman,Ravasz,NewmanGirvan,Radicchi}: their nodes are
organized into tightly-knit communities where the node-node
connections are dense, with sparser connections in-between
communities.  This structure is often hierarchically nested, with
groups of communities themselves organized into higher level modules.
Given the relevance of modularity to features like functional units in
metabolic networks~\cite{Ravasz}, it is not surprising that community
structure has become one of the most intensely studied aspects of
complex networks.  Recently, Song, Havlin and Makse~\cite{Song1,
  Song2} discovered that certain modular networks also possess another
remarkable characteristic: fractal scaling, where the hierarchy of
modules shows a self-similar nesting at all length scales.  Examples
of such fractal networks include the WWW, the actor collaboration
network, protein interaction networks in {\it E.  coli}, yeast, and
humans, the metabolic pathways in a wide variety of
organisms~\cite{Song1}, and genetic regulatory networks in {\it S.
  cerevisiae} and {\it E.  coli}~\cite{Yook}.

Despite the widespread occurrence of fractal topologies, little is yet
known about the nature of cooperative behavior on these networks, or
even more generally on how modular structure affects collective
ordering or correlations among interacting objects.  In particular the
Ising model has been investigated extensively on non-fractal
scale-free
networks~\cite{Aleksiejuk,Bianconi,Dorogov0,Leone,Igloi,Goltsev,Indekeu,Giuraniuc1,Giuraniuc2},
but only recently has a form of community structure been included: an
Ising ferromagnet was studied on two weakly coupled Barabasi-Albert
scale-free networks with a varying density of inter-network
links~\cite{Suchecki}, finding stable parallel and antiparallel
orderings of the two communities at low temperatures.  It would be
interesting to examine a system with a large number of interacting
communities, capturing more fully the complex modular organization of
real-world examples.  In this paper, we introduce a hierarchical
lattice~\cite{BerkerOstlund,KaufmanGriffiths1,KaufmanGriffiths2}
network model exhibiting a nested modular structure with fractal
scaling.  Hierarchical lattices (part of a broader class of
deterministically constructed
networks~\cite{Barabasi,Comellas,RavaszBarabasi,Andrade,Doye,Zhang1,ZhangComellas,ZhangRong,Zhang2,Zhang3})
have been the focus of increasing attention
recently~\cite{Dorogov2,HinczewskiBerker,Rozenfeld,Zhang}, since they
can be tailored to exhibit various features---including scale-free
degree distributions, small-world behavior, and fractal
structure---for which exact analytical expressions can be derived.
The explicit results from such deterministic models can serve as a
testing ground for approximate phenomenological approaches, and a
starting point for extensions incorporating additional realistic
features like randomness~\cite{Zhang}.

Here we exploit another advantage of such lattices: the ferromagnetic
Ising model can be solved through an exact renormalization-group (RG)
transformation.  Varying the strength of interactions between
communities, we find an unusual combination of thermodynamic
properties.  At high temperatures or weak inter-community coupling the
system is disordered, but the free energy as a function of magnetic
field $H$ is nonanalytic at $H=0$.  This is due to the presence of
rare large communities, similar to the Griffiths singularity in
bond-diluted ferromagnets below the percolation
threshold~\cite{Griffiths}: there the system is partitioned into
disjoint clusters of connected sites, and the small probability of
arbitrarily large clusters leads to an analogous nonanalyticity in the
free energy above $T_c$.  As we lower the temperature in our network,
true long-range order is never achieved at $T>0$, even for the
strongest inter-community coupling.  Surprisingly, we find instead a
low-temperature phase with algebraic order, just as in the $XY$ model:
the magnetization is zero, but there is power-law decay of pair
correlations with distance, and the thermodynamic functions throughout
the entire phase behave as if at a critical point.

The organization of the paper is as follows: in Sec.~II we describe
the network's construction (Sec.~II.A) and summarize its topological
properties (Sec.~II.B), including its community structure and fractal
scaling characteristics.  Sec.~III examines thermodynamic properties
of the Ising model on the network, derived from an exact
renormalization-group approach (Sec.~III.A).  We discuss the phase
diagram and critical behavior in Sec.~III.B, and then focus on two
particularly interesting aspects of the results: the presence of
Griffiths singularities in the free energy (Sec.~III.C), and the
nature of long-range pair correlations in the low-temperature phase
(Sec.~III.D).  We present our conclusions in Sec.~IV, and note that
the behavior described here is characteristic of a broader class of
hierarchical lattice complex networks---a fact that will be explored
in future studies.

\section{Network Properties}

\subsection{Construction procedure}

\begin{figure}[t]
\centering \includegraphics*{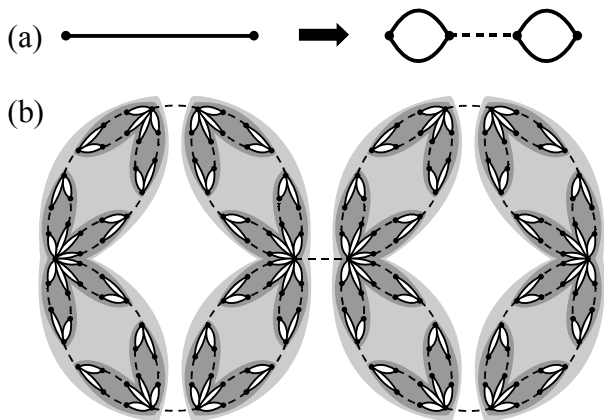}
\caption{(a) Construction of the hierarchical lattice.  (b) The
  lattice after $t=4$ steps in the construction.  Communities at the
  $n=0,1,2$ levels of the hierarchy are shown with white, dark gray,
  and light gray backgrounds respectively.}\label{fig:1}
\end{figure}

Our lattice has two types of bonds, depicted as solid and dashed lines
respectively.  At each construction step, every solid bond is replaced
by the connected cluster on the right of Fig.~\ref{fig:1}(a), and this
procedure is iterated $t$ times.  The initial $t=0$ lattice is two
sites connected by a single solid bond.  Fig.~\ref{fig:1}(b) shows the
network at $t=4$.  All results quoted below are for the infinite
lattice limit, $t \to \infty$.

\subsection{Topological characteristics}

{\it Size of network:} The total number of sites $N = \frac{2}{3}(4^t+2)$,
the total number of bonds $N_b = \frac{1}{3}(4^{t+1}-1)$, and the diameter of
the network (the maximum possible shortest-path distance between any
two sites) is $D = 2^{t+1}-1$.  

{\it Degree distribution:} The probability $P(k)$ of finding a node
with degree $k$ is zero except for $k = 2^m+1$ for some integer $m\ge
1$, where $P(k) = 3 \cdot 4^{-m}$.  The scale-free exponent $\gamma$
is calculated from the cumulative distribution $P_\text{cum}(k) \equiv
\sum_{k^\prime=k}^\infty P(k^\prime) \sim k^{1-\gamma}$.  For large
$k$ we find $P_\text{cum}(k) \approx k^{-2}$, so $\gamma = 3$.

{\it Community structure:} We can define several levels of hierarchical
modular organization, labeled by integer $n$: at the lowest level
($n=0$) we have clusters of solid bonds (shown with white background
in Fig.~\ref{fig:1}(b)), with the dashed bonds acting as
inter-community links; at the next level $(n=1)$ we can group together
those communities which correspond to a single solid-bond cluster at
the $(t-1)$th construction step (dark gray background in the figure);
the $n=2$ level communities are outlined in light gray.  In general,
for a lattice after $t$ construction steps, a community at the $n$th
level of the hierarchy evolved from a single solid-bond cluster at
step $t-n$.

{\it Fractal scaling:} Adapting the analysis in Refs.~\cite{Song1,
  Song2}, we can characterize the fractal topology of the network
through two exponents $d_B$, $d_k$, defined as follows.  At the $n$th
level, all communities have the same diameter, $\ell_B \rsp = \rsp 2^{\,n+2}-2$
(for $n < t-1$).  Thus at each level the communities form a ``box
covering'' of the entire network with boxes of the same $\ell_B$.  The
scaling of the number of boxes $N_B(\ell_B)$ required to cover the
network for a given $\ell_B$ defines the fractal dimension $d_B$,
namely $N_B(\ell_B)/N \sim \ell_B^{-d_B}$.  In our case we have
$N_B(\ell_B)/N = 4^{-n-1} \approx 4 \ell_B^{-2}$ for large
$n$, yielding $d_B = 2$.  Similarly the degree exponent $d_k$ of the
boxes is defined through $k_B(\ell_B)/k_\text{hub} \sim
\ell_B^{-d_k}$, where $k_B(\ell_B)$ is the number of outgoing links
from the box as a whole, and $k_\text{hub}$ the degree of the most
connected node inside the box.  For boxes with large $k_\text{hub}$ we
get a scaling $k_B(\ell_B)/k_\text{hub} \approx 2 \ell_B^{-1}$, giving
$d_k = 1$.  As with all the real-world fractal networks examined in
Refs.~\cite{Song1, Song2}, the scale-invariance of the probability
distribution is related to the fractal scaling of the network through
the exponent relation $\gamma = 1+d_B/d_k$, which is satisfied for
$d_B =2$, $d_k = 1$, and $\gamma=3$.

{\it Modularity:} The strength of community structure---the extent to
which nodes inside communities are more tightly knit than an
equivalent random network model---is quantified through the
modularity~\cite{NewmanGirvan} $Q = \sum_s \left[l_s/N_b -(d_s/2N_b)^2
\right]$, where the sum runs over all communities, and $l_s$, $d_s$
are the total number of bonds and total sum of node degrees for the
$s$th community.  In our case each level $n$ in the hierarchy
describes a different partition of the network into communities, and
we find the corresponding modularity $Q = 1-4^{-n-1}$.  Thus $Q$
increases from $3/4$ at $n=0$ to the maximum possible value 1 as $n
\to \infty$, showing that the modular structure becomes ever more
pronounced as we go to higher levels.

{\it Distribution of shortest-paths:} We define $N_\ell$ as the total
number of site pairs $(i,j)$ whose shortest-path distance along the
lattice $\ell_{ij} = \ell$.  The distance $\ell$ can take on values
between 1 and $D$.  $N_\ell$ has a non-trivial dependence on $\ell$,
but satisfies the scaling form $N_\ell = 2^{3t} f_t(\ell/D)$, where
the function $f_t(\ell/D)$ approaches 0 for $\ell$ close to $1$ or
$D$, and $f_t(\ell/D) \sim \text{O}(10^{-1})$ for $1\ll \ell \ll D$.
The average shortest-path length $\bar\ell =\frac{2}{N(N-1)}
\sum_{\ell=1}^D \ell N_\ell \sim D \sim N^{1/2}$, so the network is
not small-world.

\section{Ising Model on the Network}

\subsection{Renormalization-group transformation}

Let us now turn to the Hamiltonian for our system,
\begin{equation}\label{eq:1}
-\beta {\cal H}=J\sum_{\langle i j \rangle_s} s_i s_j+ K
\sum_{\langle i j \rangle_d} s_i s_j + H \sum_{i} s_i\,,
\end{equation}
where $s_i = \pm 1$, $J, K > 0$, and $\langle i j \rangle_s$, $\langle
i j \rangle_d$ denote sums over nearest-neighbor pairs on the solid
and dashed bonds respectively.  The ratio of inter- to intra-community
coupling is parametrized by $K/J$.  The RG transformation is the
reverse of the construction step: the two center sites in every
cluster like the one on the right of Fig.~\ref{fig:1}(a) are
decimated, giving an effective interaction between the two remaining
sites.  The renormalized Hamiltonian $-\beta {\cal H}^\prime$ has the
same form as Eq.~\eqref{eq:1}, but with interaction constants
$J^\prime, K^\prime, H^\prime$.  Two additional terms also appear: a
magnetic field counted along the solid bonds, $H_B^\prime
\sum_{\langle i j \rangle_s} (s_i +s_j)$, and an additive constant per
solid bond $G^\prime$.  The renormalized interaction constants are
given by:
\begin{equation}\label{eq:2}
\begin{split}
J^\prime &= (1/4)\ln\left(R_{1}R_{2}/R_{3}^2\right),\quad
K^\prime=K,\quad H^\prime=H,\\
H_B^\prime&=
(1/4)\ln\left(R_{1}/R_{2}\right),\quad G^\prime =
(1/4)\ln\left(R_{1}R_{2}R_{3}^2\right),
\end{split}
\end{equation}
where:
\begin{equation}\label{eq:3}
\begin{split}
R_{1} &=
w^{-4}xy^{-2}+2x^{-1}z^4+w^4xy^2z^8,\\
R_{2} &= w^{-4}xy^{2}+2x^{-1}
z^{-4}+w^4xy^{-2}z^{-8},\\
R_{3} &=
x^{-1}w^{-4}+x^{-1}w^4+xy^{-2}z^{-4}+xy^{2}z^{4},
\end{split}
\end{equation}
and $w = e^J$, $x=e^K$, $y=e^H$, $z=e^{H_B}$.  Under
renormalization a nonzero site magnetic field $H$ induces a bond
magnetic field $H_B$, while $H^\prime = H$ since the site field at the
edge sites in each cluster is unaffected by the decimation of the
center sites~\cite{BOP}.  This transformation is exact, preserving the
partition function $Z^\prime = Z$, and we iterate it to obtain the RG
flows, yielding the global phase diagram of the system.  Thermodynamic
densities, corresponding to averages of terms in the Hamiltonian,
transform under RG according to a conjugate recursion
relation~\cite{BOP}.  Iterating this along the flow trajectories until
a fixed point is reached, we can directly calculate the magnetization
$M = \frac{1}{N} \sum_i \langle s_i \rangle$, internal energy per site
$U = \frac{1}{N}\langle \cal H \rangle$, and their derivatives $\chi =
\frac{\partial M}{\partial H}$, $C = \frac{\partial U}{\partial T}$.

\begin{figure}
\includegraphics*{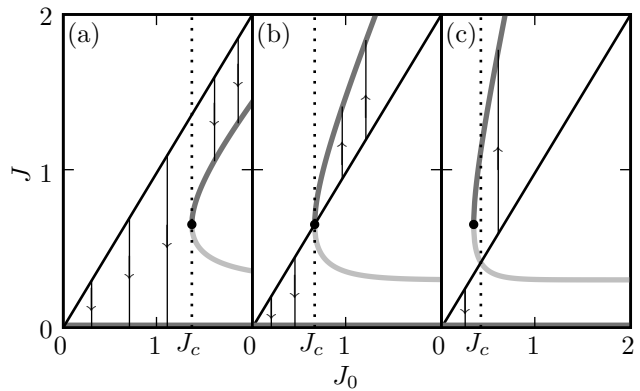}
\caption{Renormalization-group flows in the closed subspace $H=H_B=0$
  for three cases of the inter-community coupling strength $K/J$: (a)
  $K/J = 0.75$, below the threshold value $(K/J)_m =
  \ln\left(\frac{1}{11} (43+24\sqrt{3})\right)/\ln(2+\sqrt{3})\approx
  1.549$; (b) $K/J = (K/J)_m$; (c) $K/J = 3 > (K/J)_m$.  In each case, the
  diagonal straight line represents the initial condition $J=J_0 =
  1/T$, the thin vertical lines with arrows are sample flows, and the
  thick gray lines represent fixed points, with dark gray
  corresponding to stable fixed points, and light gray corresponding
  to unstable fixed points.  The point where the light and gray curves
  meet is marginally stable.  The dashed line marks the critical $J_c$
  separating the disordered phase, flowing to the phase sink $J^\ast
  =0$, and the algebraically ordered phase, flowing to a line of
  finite temperature fixed points $J^\ast(J_0)$ (the dark gray curve).}
\label{flows}
\end{figure}

We are also interested in the pair correlation $G_{ij} = \langle s_i
s_j \rangle - \langle s_i \rangle \langle s_j \rangle$ for arbitrary
sites $i$, $j$ in the network.  Since our lattice is highly
inhomogeneous, $G_{ij}$ is not a simple function of $\ell_{ij}$.
However, following the analysis in Ref.~\cite{Dorogov1}, we can define
an average correlation $G(\ell) = \frac{1}{N_\ell}\sum_{\{(i,j) :
  \ell_{ij} = \ell\}} G_{ij}$, where the sum is over all pairs $(i,j)$
satisfying $\ell_{ij} = \ell$.  While $G(\ell)$ cannot be directly
calculated from the RG flows, we can determine its long-distance
scaling properties.  Moreover, the average correlation for a certain
subset of pairs in the lattice can be explicitly calculated: at the
$n$th level of the hierarchy, let site $i$ be a hub of a community,
and $j$ be a site at the very edge of the same community, separated by
a distance $\ell_n=2^{n+1}-1$.  Denote the average of $G_{ij}$
restricted to this subset as $G_\text{hub}(\ell_n)$.  After $n$ RG
steps, such pairs $(i,j)$ become nearest-neighbors along a solid bond,
and thus we can obtain their thermodynamic average through the
conjugate recursion relation~\cite{BOP}.  The number of such pairs is
$2^{2t-2n-1}$.  For $1 \ll \ell_n \ll D$, compared to the overall
number of pairs with the same separation, $N_{\ell_n} \sim 2^{3t}$,
the subset forms a vanishingly small fraction of the total in the $t
\to \infty$ limit.

\begin{figure}
\includegraphics*{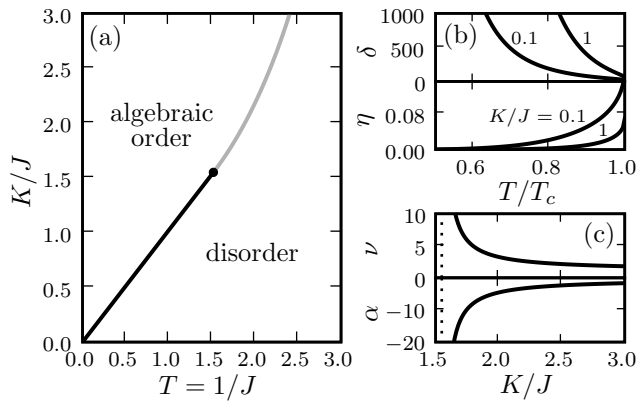}
\caption{(a) Phase diagram.  The black and gray curves represent
  infinite- and second-order transitions respectively. (b,c) Critical
  exponents.  The dotted line marks $(K/J)_m$.}\label{phase}
\end{figure}

\subsection{Phase diagram and critical properties}

Fig.~\ref{flows} depicts various cases for the renormalization-group
flows, and the corresponding phase diagram in terms of temperature
$T=1/J$ versus $K/J$ at $H=0$ is shown in Fig.~\ref{phase}(a).  The
two phases, both with $M=0$, are (1) a disordered phase where pair
correlations decay exponentially, $G(\ell) \sim \exp(-\ell/\xi)$ with
a finite correlation length $\xi$; (2) a phase with algebraic order,
$\xi = \infty$, characterized by power-law decay of correlations,
$G(\ell) \sim \ell^{-\eta(T,K/J)}$.  Since this latter phase flows
under RG to a line of finite-temperature fixed points (the dark gray
curves in Fig.~\ref{flows}), a different fixed point for every value
of $T$ and $K/J$, we have a varying exponent $\eta(T,K/J)$.
Fig.~\ref{phase}(b) plots $\eta$ for $K/J = 0.1$ and $1$, and we note
that $\eta \to 0$ for $T \to 0$, as the system asymptotically
approaches true long range order at $T=0$.  Strengthening
inter-community coupling has a similar effect, with $\eta$ decreasing
for larger $K/J$.

For $K/J$ below a threshold value $(K/J)_m \equiv
\ln\left(\frac{1}{11} (43+24\sqrt{3})\right)/\ln(2+\sqrt{3}) \approx
1.549$, the phase transition is infinite order: as $T \to T_c^+$ we
have exponential singularities of the
Berezinskii-Kosterlitz-Thouless~\cite{Berezinskii,KosterlitzThouless}
(BKT) form, just like in the $XY$ model.  $\xi \sim e^{A/\sqrt{t}}$
and the singular part of the specific heat $C_\text{sing} \sim
e^{-B/\sqrt{t}}$, where $t \equiv (T-T_c)/T_c$ and the constants
$A,B>0$.  It is interesting to note that for $K/J =1$, our Hamiltonian
can be mapped by duality transformation to the Ising model on the
small-world hierarchical lattice of Ref.~\cite{HinczewskiBerker} (the
$p=1$ lattice).  On this dual network a similar BKT transition occurs,
though with the algebraic order in the high temperature phase (much as
the Villain version of the $XY$ model is dual to a discrete Gaussian
model describing roughening, with the low and high-temperature
properties reversed \cite{ChuiWeeks}).  BKT singularities have also
been observed in an Ising and Potts system on an inhomogeneous growing
network~\cite{Bauer,Khajeh}.  For $K/J > (K/J)_m$, on the other hand,
the phase transition is second-order: $\xi \sim t^{-\nu}$,
$C_\text{sing} \sim t^{-\alpha}$ for exponents $\alpha$, $\nu$,
plotted as a function of $K/J$ in Fig.~\ref{phase}(c).  Thus with
increasing $K/J$ the system almost looks like an ordinary second-order
Ising transition: a critical $T_c$ below which we have something very
close to long-range order, since $\eta$ is nearly zero.

\begin{figure}
\includegraphics*{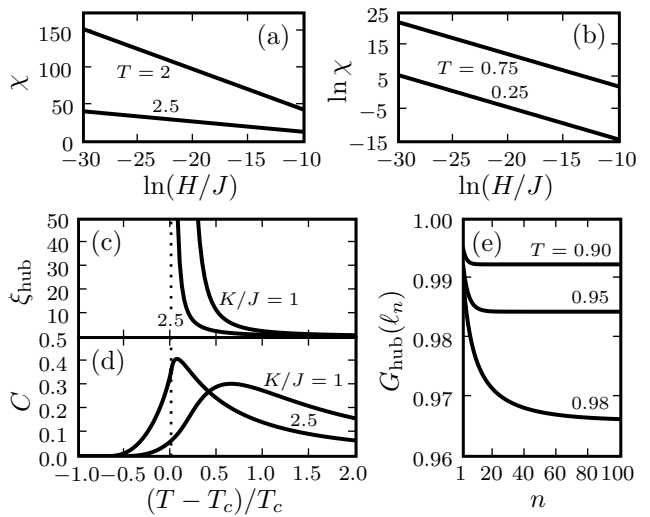}
\caption{(a) Magnetic susceptibility $\chi$ at $K/J = 1$ for $T =
  2,\,2.5 > T_c$.  (b) $\chi$ at $K/J=1$ for $T=0.75,\,0.25 < T_c$.
  (c,d) Hub correlation length $\xi_\text{hub}$ and specific heat $C$
  at $K/J = 1,\,2.5$.  (d) Hub correlation function
  $G_\text{hub}(\ell_n)$, where $\ell_n = 2^{n+1}-1$, at $K/J = 1$ for
  $T = 0.90,\,0.95,\,0.98 < T_c$.}\label{dens}
\end{figure}

\subsection{Griffiths singularities}

The disordered phase in our network differs in one important aspect
from a conventional paramagnetic phase: $M \sim H(1-\ln H)$ for small
$H$, leading to a divergence in the susceptibility, $\chi \sim -\ln
H$.  We see this in Fig.~\ref{dens}(a) for $T = 2$, $2.5$, and $K/J =
1$.  As mentioned above, the mechanism for this nonanalyticity at
$H=0$ is similar to the one behind the Griffiths singularity in random
ferromagnets.  We can understand it through the distribution of
Yang-Lee zeros~\cite{YangLee} of the partition function in the complex
magnetic field plane.  Introducing the variable $z = e^{-2H}$, the
Yang-Lee theorem states that the zeros of $Z$ lie on the unit circle
$z=e^{i\theta}$ in the complex $z$ plane.  If $g(\theta)$ is the
density of these zeros (a continuous distribution in the thermodynamic
limit), then for a regular ferromagnet above $T_c$ we have $g(\theta)
= 0$ for a finite range of $\theta$ near the real axis $\theta=0$, so
that $Z$ is analytic at $H=0$.  However in our case $g(\theta)$
``pinches'' the real axis: $g(0)=0$ but $g(\theta \ne 0) > 0$.  Since
$g(\theta)$ is related to the magnetization through $g(\theta) =
\frac{1}{2\pi} \lim_{r \to 1^-} \text{Re}\, M(z=re^{i\theta})$, we can
deduce from the observed singularity that $g(\theta)\sim \theta$ for
small $\theta$.  The dominant contributions to this $g(\theta)$ come
from large communities, centered at hubs with degree $k=2^m+1$ for $m
\gg 1$, despite their small probability $P(k) =3\cdot 4^{-m}$. 

We can see these contributions explicitly for the $K/J = 0$ case,
adapting arguments used to derive scaling forms for $g(\theta)$ in
disordered ferromagnets~\cite{BrayHuifang,Chan}.  The system in this
case is a disjoint set of solid-bond clusters, with the probability of
a randomly chosen site being part of a cluster of size $N_m = 2^{m-1}
+1$ given by $P_m = 3(2+2^m)/2^{2m+1}$ (for $m\ge 2$).  The average
magnetization per site of such of cluster is easily calculated
analytically, and takes the following approximate form for small $H$,
\begin{equation}\label{eq:4}
M_m \approx \tanh(H f_m(J))\,,
\end{equation}
where
\begin{equation}\label{eq:5}
\begin{split}
&f_m(J)\\
&= 1+\frac{N_m-1}{N_m}\tanh(2J)\left(2+(N_m-2)\tanh(2J)\right)\,.
\end{split}
\end{equation}
The $H f_m(J)$ term in Eq.~\eqref{eq:4} can be interpreted as the
effective field felt by the cluster, with the function $f_m(J)$
varying between $1$ at $J=0$ and $N_m$ at $J=\infty$.  In the large
$m$ limit $f_m(J) \to N_m \tanh^2(2J)$ for all $J>0$.  To find
$g_m(\theta)$, the cluster's contribution to the overall $g(\theta)$,
we plug in a small complex magnetic field $H =
\frac{1}{2}(\epsilon-i\theta)$, corresponding to $z =
(1-\epsilon)e^{i\theta}$, and take the real part of the resulting
magnetization: $g_m(\theta) = \frac{1}{2\pi}\lim_{\epsilon\to 0}
\text{Re}\,M_m(H=\frac{1}{2}(\epsilon-i\theta))$.  This gives
\begin{equation}\label{eq:6}
g_m(\theta) = \lim_{\epsilon\to 0} \frac{1}{2\pi} \frac{\sinh(\epsilon
  f_m(J))}{\cos(\theta f_m(J))+\cosh(\epsilon f_m(J))}\,.
\end{equation}
This expression is dominated by high, narrow peaks at $\theta =
(2n+1)\pi/f_m(J)$, $n=0,1,\ldots$, and we can write it as a sum of
delta functions in the small $\epsilon$ limit,
\begin{equation}\label{eq:7}
\begin{split}
  g_m(\theta) &\approx \lim_{\epsilon\to 0} \frac{1}{\pi} \sum_{n=0}^{\infty}
  \frac{\epsilon f_m(J)}{f_m^2(J)\left(\theta-\frac{(2n+1)\pi}{
    f_m(J)}\right)^2+\epsilon^2
    f_m^2(J)}\\
  &= \sum_{n=0}^{\infty} \frac{1}{f_m(J)}\, \delta\left(\theta-\frac{(2n+1)\pi}{f_m(J)}\right)\,.
\end{split}
\end{equation}
Thus the total $g(\theta)$ for the system is
\begin{equation}\label{eq:8}
g(\theta) = \sum_{m=2}^{\infty} \sum_{n=0}^{\infty}  \frac{P_m}{f_m(J)}\, \delta\left(\theta-\frac{(2n+1)\pi}{f_m(J)}\right)\,.
\end{equation}
For small $\theta$, the nonzero contributions to $g(\theta)$ come from
$m$ values where $f_m(J) = (2n+1)\pi/\theta$ for some $n$.  Since
$f_m(J) \approx N_m \tanh^2(2J)$ for large $m$, these are the
contributions of clusters with large size $N_m \propto 1/\theta$, with
a corresponding probability $P_m \approx 3/4N_m$. Eq.~\eqref{eq:8}
becomes
\begin{equation}\label{eq:9}
  g(\theta) \approx \sum_{m,n}  \frac{3\theta^2 \tanh^2(2J)}{4(2n+1)^2 \pi^2}
  \, \delta\left(\theta-\frac{(2n+1)\pi}{N_m \tanh^2(2J)}\right)\,.
\end{equation}
As $m\to \infty$ the delta function peaks become densely spaced, and
from Eq.~\eqref{eq:9} it is evident for small $\theta$ that
$g(\theta)$ scales like $g(b \theta) = b g(\theta)$ for any constant
$b$, consistent with the observation of $g(\theta) \sim \theta$
deduced from the singularity in $M$.  Thus we see this behavior is
directly related to the presence of large communities around highly
connected hubs, which have a scale-free distribution $P_m \sim
N_m^{-1}$.

In comparison, for bond-diluted ferromagnets below the percolation
threshold large connected clusters are exponentially rare, the
resulting $g(\theta) \sim e^{-A(T)/|\theta|}$, and the Griffiths
singularity is much weaker, leading to a finite $\chi$ at
$H=0$~\cite{Harris}.  Turning now to the algebraic phase for $T <
T_c$, here $M$ and $\chi$ behave as if at a critical point: $M \sim
H^{1/\delta(T,K/J)}$, $\chi \sim H^{1/\delta(T,K/J)-1}$ as $H \to 0$.
Fig.~\ref{dens}(b) shows $\chi$ for $K/J =1$, $T=0.75$, $0.25$, and
Fig.~\ref{phase}(b) plots the exponent $\delta(T,K/J)$ for $K/J = 0.1$
and $1$.  The corresponding scaling of the density of zeros is
$g(\theta) \sim \theta^{1/\delta(T,K/J)}$ near $\theta=0$.
\vspace{1em}

\subsection{Pair correlations}

Finally, we consider the behavior of the hub correlation function
$G_\text{hub}(\ell_n)$.  For $T>T_c$, we find an exponential decay
with $\ell_n$, which we characterize by a correlation length
$\xi_\text{hub}$.  The divergence of $\xi_\text{hub}$ as $T \to T_c^+$
is shown in Fig.~\ref{dens}(c) for $K/J = 1$, $2.5$. Like the overall
pair correlation length $\xi$, $\xi_\text{hub}$ diverges with a BKT
form for $K/J<(K/J)_m$ and as a power law for $K/J>(K/J)_m$.  The
onset of the rapid increase in $\xi_\text{hub}$ coincides with the
position of the peak in the specific heat $C$, plotted in
Fig.~\ref{dens}(d).  Just like in the $XY$ model~\cite{BerkerNelson},
$C$ is smooth at $T_c$ for all $K/J$, and the peak occurs at $T>T_c$,
corresponding to the onset of short-range order in the system.  For
$T<T_c$, $G_\text{hub}(\ell_n)$ has a surprising behavior: as seen in
Fig.~\ref{dens}(e), it approaches a nonzero limit as $\ell_n \to
\infty$, a signature of long-range order.  However, since
$G_\text{hub}(\ell_n)$ describes only a subset of pairs, a vanishingly
small fraction of the total for large $\ell_n$, the long-range
ordering of these pairs is compatible with $M$ being zero.  The
presence of such long correlations in the algebraic phase, and the
overall slow power-law decay of $G(\ell)$ with $\ell$, is remarkable
given that for ``fat-tailed'' scale-free networks (i.e. with $\gamma
\le 3$) pair correlations longer than nearest-neighbors are typically
suppressed: one can prove that $G(\ell>1) = 0$ at $H=0$ in the
thermodynamic limit, if $\chi(H=0)$ is finite~\cite{Dorogov1}.  In our
case $\chi(H=0) = \infty$ at all $T$, the proof does not apply, and we
see that this fractal modular lattice is an important exception to the
general expectation of weakened pair correlations on
networks~\cite{Dorogov1}.

\section{Conclusions}

In conclusion, we have introduced a hierarchical lattice network with
the modular structure and fractal scaling characteristic of a wide
array of real-world networks.  The Ising model on this
lattice---solved through an exact RG transformation---exhibits an
interesting transition.  A disordered phase with Griffiths
singularities gives way at low temperatures to algebraic order: the
system behaves as if at criticality for a broad range of parameters,
and we find power-law decay of pair correlations, unexpected for this
type of scale-free network.

The thermodynamic phenomena observed here are not confined to one
particular network.  In fact, we can consider the much larger class of
hierarchical lattices that form scale-free networks on which the Ising
model exhibits a standard order-disorder transition: these include
fractal lattices on which Migdal-Kadanoff recursion relations are
exact~\cite{Migdal,Kadanoff}, related hybrid lattices~\cite{Erbas},
and their duals~\cite{HinczewskiBerker2}.  In all these cases we can
modify the connected graph which defines the lattice construction step
as follows: replace a subset of the bonds in the graph with dashed
bonds (which remain unaltered as we iterate the construction) in such
a way that the graph would break into two or more disjoint pieces if
the dashed bonds were cut.  The Ising model on the resulting
hierarchical lattice will no longer flow under renormalization to an
ordered fixed point at low temperatures, but rather to a continuous
line of fixed points, yielding an algebraically ordered phase.  And
the power-law distribution of highly connected hubs in such networks
will lead to Griffiths singularities in the disordered phase.
Conversely the duals of such networks---which have infinite fractal
dimension and show small-world scaling---have algebraic order at high
temperatures.  Recent studies have highlighted the diversity of
structural properties in families of hierarchical lattice
networks~\cite{Rozenfeld,Zhang}---the ability to tune degree
exponents, fractal dimensionality, and other topological aspects of
these networks by varying the defining graph.  The structural richness
of these networks is manifested through unusual phase transitions and
critical phenomena, already apparent even in a simple system like the
Ising model.  The use of renormalization-group methods to characterize
cooperative behavior on this broad class of networks, both in the pure
case and in the presence of bond randomness~\cite{HinczewskiBerker},
will be the subject of future work.

I thank A.N. Berker, T. Garel, and H. Orland for useful discussions.

\end{document}